\documentclass[conference,a4paper]{IEEEtran}

\usepackage{graphicx, caption, subcaption}
\usepackage{algorithm}
\usepackage{array}

\usepackage{algorithmicx}
\usepackage{algpseudocode}

\usepackage{soul}
\usepackage{url}
\usepackage{verbatim}
\usepackage{cite}
\usepackage{listings}
\usepackage[table]{xcolor}
\usepackage[nolist]{acronym}
\usepackage{amsmath}
\usepackage{amssymb}
\usepackage{soul}
\usepackage{rotating}
\usepackage{multirow} 

\usepackage{mathtools, stmaryrd}
\usepackage{xparse} \DeclarePairedDelimiterX{\Iintv}[1]{\llbracket}{\rrbracket}{\iintvargs{#1}}
\NewDocumentCommand{\iintvargs}{>{\SplitArgument{1}{,}}m}
{\iintvargsaux#1} %
\NewDocumentCommand{\iintvargsaux}{mm} {#1\mkern1.5mu..\mkern1.5mu#2}

\usepackage{siunitx}

\usepackage{blindtext}
\usepackage{ctable} 
\def\BibTeX{{\rm B\kern-.05em{\sc i\kern-.025em b}\kern-.08em
    T\kern-.1667em\lower.7ex\hbox{E}\kern-.125emX}}

\newcommand\Tstrut{\rule{0pt}{2.6ex}}         
\newcommand\Bstrut{\rule[-0.9ex]{0pt}{0pt}}   

\begin{document}

\voffset=0.05in
\textheight=9.28in

\title {Effective ML Model Versioning in Edge Networks}
\author{\IEEEauthorblockN{Fin Gentzen, Mounir Bensalem and Admela Jukan}
\IEEEauthorblockA{Technische Universit\"at Braunschweig, Germany;
\{f.gentzen, mounir.bensalem, a.jukan\}@tu-bs.de}

}

\maketitle

\begin{abstract}
Machine learning (ML) models, data and software need to be regularly updated whenever essential version updates are released and feasible for integration. This is a basic but most challenging requirement to satisfy in the edge, due to the various system constraints and the major impact that an update can have on robustness and stability. In this paper, we formulate for the first time the ML model versioning optimization problem, and propose effective solutions, including the update automation with reinforcement learning (RL) based algorithm. We study the edge network environment due to the known constraints in performance, response time, security, and reliability, which make updates especially challenging. The performance study shows that model version updates can be fully and effectively automated with reinforcement learning method. We show that for every range of server load values, the proper versioning can be found that improves security, reliability and/or ML model accuracy,  while assuring a comparably lower response time. \end{abstract}


\section{Introduction}

The field of AI and ML has gone through unprecedented advancements over the past decade, whereby AI/ML models are continuously improving in size, performance, robustness and accuracy \cite{paleyes2022challenges}.   In this rapidly evolving field, ML-based domain applications, such as in industrial engineering and health care, need to be continuously updated, along with the related ML software updates as well as data updates. To assure stability and manage applications and systems that use ML, an effective versioning of ML models, data and code is critical. Considering that most network management operations rely on ML based models, and most applications in IoT, cloud and edge computing today require ML-based functions, this challenge  appears even more significant. Recent ML Operations (MLOps) efforts also recognize that automating and operationalizing ML products presents a grand challenge \cite{kreuzberger2023machine}.

\par Model robustness is one of the key challenges in ML today, which requires creation of new versions of ML models \cite{angioni2024robustness}. The ML model version updates, on the other hand, add management overhead, and in general can decrease the stability of the system. Adding any new and enhanced model is contingent upon modifying the existing system, and any changes fundamentally carry risks to destabilize the performance \cite{krishnaswamy2022decentralized}. In some domain applications, like health care, stability is premium, due to safety \cite{10.1145/3450439.3451864}.   When updating ML models in edge networks with constrained resources, several conflicting objectives need to be considered, such as accuracy, reliability, security, system stability, and resource utilization \cite{45305}.  Only a few papers so far addressed this issue, like in \cite{10597011} on how to update ML model for data series prediction, focused on a a single ML model. Today, multiple ML models updates need to be considered with various conflicting constraints. 

In this paper, we focus on the problem of ML model versioning and updates in edge networks. To this end, we formulate for the first time the ML versioning optimization problem and propose a reinforcement learning (RL)-based decision making algorithm to automate and optimize the ML model update process. We also propose the related functional network management architecture of an \textit{Update Manager}. We analyze the performance of our proposed approach by comparing it to conventional update methods used in today's systems.  The results show that under moderate levels of traffic in the network,  our algorithm achieves a faster response times as compared to the manual policy to always update when there is a new version available. It also significantly improves ML model parameters, such as accuracy, security and reliability. With high server loads, the automation with RL is less beneficial, due to the fact that in a system under stress, the request queues become full and model replicas remain alive during the runtime. In that case, the additional delay by an update slows down the request processing in the queue.

\par The rest of the paper is organized as follows. 
Section \ref{sec:model} describes the reference architecture. Section \ref{sec: problem} formulates the ML model versioning optimization problem.  Section \ref{sec:solution} describes the RL-based solution.  Section \ref{sec:results} presents numerical results. Section \ref{sec:conclusion} concludes the paper.

\begin{figure*}
 \centering 
   \includegraphics[scale=0.55]{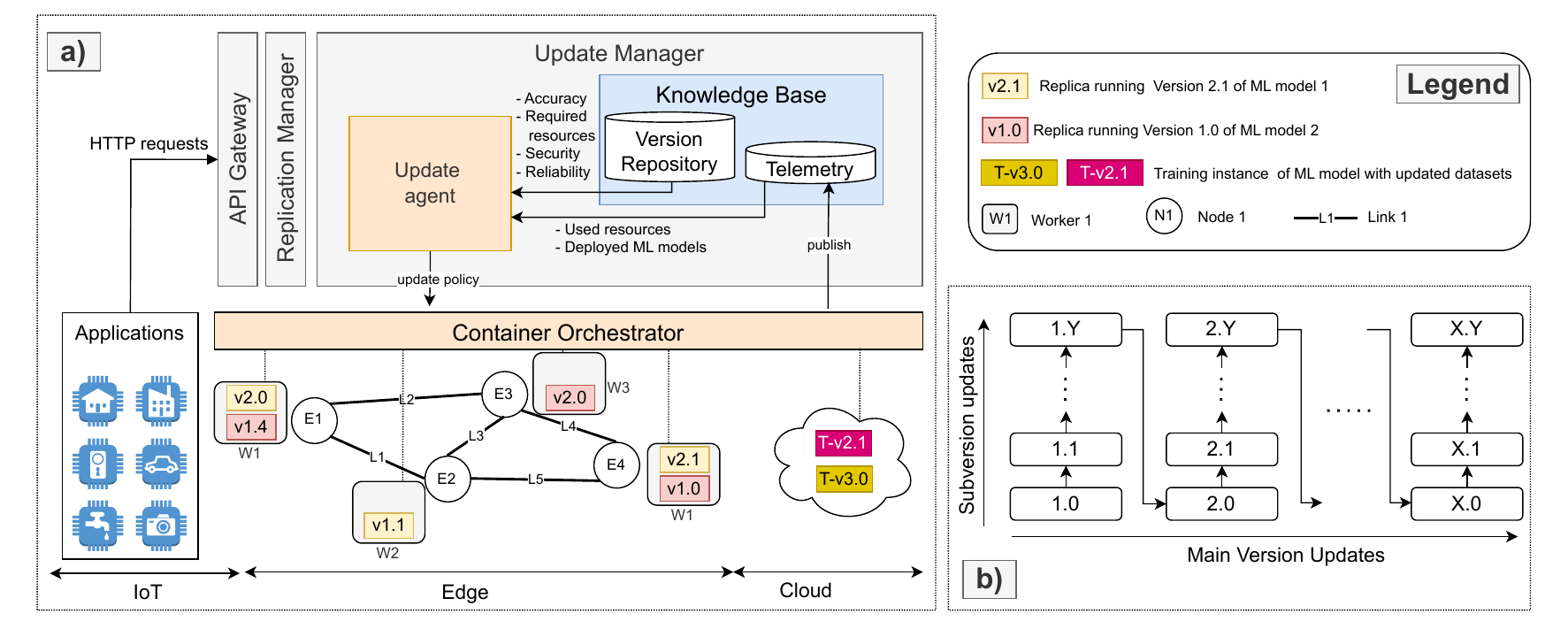}
 \caption{ a) Reference architecture, b)  Versions representation of an ML model.}
\label{fig:system}
\end{figure*}
\section{Reference Architecture}\label{sec:model}

Fig. \ref{fig:system}.a) depicts the reference system architecture. The sample physical network consists of four edge computing nodes (E1, E2, E3 and E4) connected by links (L1, L2, L3, L4 and L5), along with the related worker machines (W1, W2, W3 and W4),  similar to our
previous work \cite{9145449}.  We assume that a system also includes a container orchestrator, typically Kubernetes, an API gateway, a replication manager, and an \emph{Update Manager} (our proposal). The system can run multiple IoT-based domain applications, as shown on the left hand side, such as vehicular, smart city, etc. The ML training is assumed in the cloud. We envision the entire system as an example of what is today known as IoT-edge-cloud continuum. Applications use  API Gateway to communicate with the system via the HTTP requests.

\par  Let us assume that all applications here are the actual ML applications, and that for each ML application, i.e., ML model, the replication manager decides the creation and removal of model replicas as well as the queuing of the received  HTTP requests towards the model updates. The nodes in the edge network are assumed to be edge devices that can host a number of dockerized containers, depending on the capacity. Each of these containers belongs to a cluster that is managed by the container orchestrator. As previously noted, the new versions of machine learning models are assumed to be  trained in the cloud and subsequently received by the container orchestrator; training is out of scope of this paper.

\par We propose the architecture of \textit{Update Manager} module, which in order to work, consists of a \textit{Telemetry} tool, a \textit{Knowledge Base}, and, in our proposal, an \textit{Update Agent}.  Telemetry is required to store the logs of all information coming from the container orchestrator, such as topology information and information about newly trained ML models. Knowledge Base is required to include a  Version Repository database.  Version Repository is expected to maintain a record of detailed information belonging to the various ML model versions, including data on the resources required, the reliability and security.  Update Agent run algorithms that decide whether to update the existing ML model. It uses the information about newly available versions from the Version Repository database and the information regarding the current state of the topology from the Telemetry module to make update decisions. Whereas \textit{Replication Manager} takes scaling decisions by deciding to deploy a new replica of a certain ML model,  \textit{Update Manager} decides  whether to utilize the oldest, most stable version, or the most recently developed version of an ML model. In the latter case, the update agent will examine the already deployed model replicas. After executing a request, it will check the availability of a new version and determine whether the existing replica should be used to process the next requests or  should be terminated to spawn a new replica with the newest version available. The update agent can use optimization algorithms, greedy heuristics, or it can be trained to find the optimal policy for the decision process.

Fig. \ref{fig:system}.b) illustrates the ML model versioning. A version is composed of a main version $X$ and a subversion $Y$, i.e., $X.Y$. The main version update replaces the entire ML model, whereas a subversion update only updates specific features, such as security. We assume that a main ML version update must be scheduled at a fixed time as it changes all model features. Subversion updates can, on the other hand, occur at random times. In this paper, we consider three equally likely, and practically relevant types of subversion updates: Security, Reliability and Accuracy updates.

\section{ML Model Versioning Problem Formulation} \label{sec: problem}
\subsection{Assumptions}
The notation is provided in Table \ref{tab:notations}. We consider a single master multi-worker deployment, including a set of edge nodes $\epsilon = \{ E_1, \ldots, E_n, \ldots, E_N \}$, where $E_n$ represents the $n^{\text{th}}$ edge node. We assume that each edge node is constrained by a certain amount of capacity modeled
as a number of CPU units $C_n$, RAM space $R_n$ and Disk space $D_n$. We consider a set of ML models denoted as $\kappa = \{ 1, \ldots, k, \ldots, K \}$. We assume that each ML model $k$, can have a main version $x\in [0, X]$, and a subversion $y\in [0,Y]$. We define a decision variable $a=1$ when  updating a ML model $k$ of version $x.y$ to a newer   version $[x+1].y$ or $x.[y+1]$, 0 otherwise. We define a mapping function $g(x,y,a)$ that gives the new version of an ML model $k$ of version $x.y$ after a decision update $a$. Each version $x.y$ of a ML model $k$ requires an amount of resources $b_{k,x,y} = [b_{k,x,y}^\text{cpu}, b_{k,x,y}^\text{ram}, b_{k,x,y}^\text{disk}]$, of CPU units, RAM and disk space.
Each ML model $k$ of version $x.y$ has a average service rate $\mu_{k,x,y}$ and an average inter-arrival rate $\lambda_{k,x,y}$, assuming for simplification that $\mu_{k,x,y}=\mu, \lambda_{k,x,y}=\lambda,$  $ \forall k,x,y$ . We adopt an $M/M/N$ queuing system to analyze the system behavior and to compute the server load (in Erlang):

\begin{equation}
    \label{eq:load}
    \text{load} = \frac{\lambda}{N \cdot \mu}
\end{equation}

\subsubsection{ML Model Placement}
Each ML model of class k requires an amount of resource that
can be served by at least one available edge node, such that:
\begin{equation}\label{eq:cpub}
    b_{k,x,y}^\text{cpu} \leq \max_{ \forall n \in [1, N]} \{ C_n \}, \quad \forall k,x,y \in [1, K] \times [0,X] \times  [0,Y]
\end{equation}
\begin{equation}\label{eq:ramb}
    b_{k,x,y}^\text{ram} \leq \max_{\forall n \in [1, N]} \{ R_n \}, \quad \forall k,x,y \in [1, K] \times [0,X] \times  [0,Y]
\end{equation}
\begin{equation}\label{eq:diskb}
    b_{k,x,y}^\text{disk} \leq \max_{\forall n \in [1, N]} \{ D_n \}, \quad \forall k,x,y \in [1, K] \times [0,X] \times  [0,Y]
\end{equation}

We denote by $\delta_{k,x,y}(n)$ the number of replicas from ML model $k$ of version $x.y$ allocated in node $E_n$. The resource allocation has
the capacity constraints, i.e.,

\begin{equation}\label{eq:cpus}
    \sum_{k=1}^{K} \sum_{x=1}^{X} \sum_{y=1}^{Y} b_{k,x,y}^\text{cpu} \delta_{k,x,y}(n) \leq C_n, \quad \forall n \in [1, N]
\end{equation}
\begin{equation}\label{eq:rams}
    \sum_{k=1}^{K}\sum_{x=1}^{X} \sum_{y=1}^{Y} b_{k,x,y}^\text{ram} \delta_{k,x,y}(n) \leq R_n, \quad \forall n \in [1, N]
\end{equation}
\begin{equation}\label{eq:disks}
    \sum_{k=1}^{K}\sum_{x=1}^{X} \sum_{y=1}^{Y} b_{k,x,y}^\text{disk} \delta_{k,x,y}(n) \leq D_n, \quad \forall n \in [1, N]
\end{equation}

\subsubsection{Delays} 
We consider four types of delays: processing delay, transmission delay, delay to spawn a replica (where applicable) and queuing delay. 
The processing delay $\tau^p_{k,x,y,n}$ of an ML model
$k$ of version $x.y$ deployed through a pod in node $E_n$ is distributed around an average value. We consider transmission and propagation times to be a constant parameter since they depend on the distance between worker (edge) nodes and the so-called master node (corresponding to the practical case in Kubernetes).
In a single master multi-worker deployment, the routing and path computation
is typically managed by the container orchestrator (e.g., Kubernetes). 
We denote by $\tau^t_{k,n}$ the transmission delay of a request $f(k)$ between master node and worker node $E_n$. 
The spawn time $\tau^s_{k,n}(a)$ of a replica of ML model $k$ of version $x.y$ with a decision update $a$ is assumed to be constant for all types of ML models. The creation of a new replica is triggered by either the Replication Manager while scaling up the system, or by  Update Manager, when we decide to update the deployed replica of the ML model $k$ with version $x.y$ to a newer version $x.[y+1]$ or $[x+1].y$. When the ML model $k$ is not updated, we set the spawn time  $\tau^s_{k,n}(a=0)=0$.

We assume the system has a queuing buffer for each ML model $k$, assessed by Replication Manager, which can use a smart monitoring algorithm, such as the described in \cite{10437794}, to scale the system. 
We denote by $\tau^q_{f(k)}$  the total queuing delay of a request  $f(k)$ of  ML model $k$.  
The queuing delay is measured as the difference between  the departure request time and the arrival request time plus the transmission plus the spawn time and the processing time. Finally, the total delay $\tau_{f(k)}$ for each  request $f(k)$  of ML model $k$, processed using a replica that runs version $x.y$ is given as:
\begin{equation}\begin{split}
    \tau_{f(k)}(x,y, a) =& \tau^p_{k,g(x,y,a),n}+ \tau^t_{k,n} + \tau^s_{k,n}(a)  
    \\ &+ \tau^q_{f(k)}, \quad  \forall k \in [1, K]
    \end{split}
\end{equation}
The average  delay for all the requests, defined as an objective to minimize, is given by:
\begin{equation}
    \mathcal{O}_1(\Delta, a) = \frac{1}{KF} \sum_{k=1}^{K}\sum_{f=1}^{F} \tau_{f(k)} (x,y, a)
\end{equation}
where $\Delta$ describes the state of the network, including nodes, and deployed replicas of ML models and their versions.
\subsubsection{ML Model Accuracy}
After training an ML model $k$, we obtain a new version that is assumed to have a higher accuracy.  We denote by  $Acc_{k}$($x,y, a$) the accuracy of a ML model $k$ with version $x.y$ after taking an update decision $a$. The average accuracy of  all requests is given by:
\begin{equation}
    \mathcal{O}_2(\Delta, a) = \frac{1}{KF} \sum_{k=1}^{K}\sum_{f=1}^{F} Acc_{k}(x,y, a)
\end{equation}

\subsubsection{ Security and Reliability}
We assume that updates can improve the security and reliability of an ML model, defined as parameters $Sec_{k}(x,y, a), Rel_{k}(x,y,a)$, respectively, assigned to ML model $k$ with version $x.y$ after taking an update decision $a$. We define two objective functions for the security and reliability as an  average among all requests  given by:
\begin{equation}\begin{split}
    \mathcal{O}_3(\Delta, a) = \frac{1}{KF} \sum_{k=1}^{K}\sum_{f=1}^{F} Sec_{k}(x,y, a)\\
    \mathcal{O}_4(\Delta, a) = \frac{1}{KF} \sum_{k=1}^{K}\sum_{f=1}^{F} Rel_{k}(x,y, a)
    \end{split}
\end{equation}

\subsection{Problem Formulation}
We define our ML Model Versioning Problem (MMV)  as a multi-objective optimization  problem, aiming at maximizing the average ML model accuracy, security and reliability of all processed requests while minimizing the average delay. The MMV problem is formulated as follows:
\begin{equation}\begin{split}
    \max_{a\in \mathcal{A} } & \left( -  \mathcal{O}_1(\Delta, a), \mathcal{O}_2(\Delta, a), \mathcal{O}_3(\Delta, a), \mathcal{O}_4(\Delta, a)  \right)  \\
    \text{subject to:   }& {Eq.} (\ref{eq:cpub}), (\ref{eq:ramb}), (\ref{eq:diskb}), (\ref{eq:cpus}), (\ref{eq:rams}), (\ref{eq:disks})
\end{split}
\end{equation}

\begin{table}[h!]
  \begin{center}
    \caption{Notations}
    \label{tab:notations}
    \begin{tabular}{c|l}
    \hline 
      \textbf{Notation} & \textbf{Description} \Tstrut\Bstrut \\
      \hline
     $N$, $L$ &   Number of edge nodes and links \Tstrut \\
      $C_n$, $R_n$, $D_n$ & N. of CPU cores, RAM and disc  in node $n$ \\
      $K$ &   Number of ML models. \\
      $f(k)$ & Request to ML model $k$\\
      $b_{k,x,y}$ & Required resources by ML model $k$\\
      $X$ & Number of main versions of an ML model \\
      $Y$ & Number of subversion of  ML model \\
      $\delta_{k,x,y}(n)$ & Number of replicas of ML model $k$ \\ & with version $x.y$ in node $n$ \\
      $\tau^{f(k),k}$ & Total delay of $f(k)$\\ 
      $\tau^p_{k,g(x,y,a),n}, \tau^t_{k,n}, $ & Processing, transmission, spawning, and queuing\\ $\tau^s_{k,n}(a), \tau^q_{f(k)}$& delay  of $k$/$f(k)$ in node $n$, respectively \\ 
     %
      $\mu$,  $\lambda$ & Average service and arrival rates\\
      
      $S$ & Set of states representation for RL \\
      $Z_k$ & Set of resource availabilities  for ML model  $k$\\
      $f_k$ & Integer variable that defines the ML model $k$\\
      $Q_k$ & Queue size of queue for ML model $k$\\
      $e$ & Binary values to indicate the type of event\\
      $\Omega$ & Set of binary values indicating the type of update\\
      $a(s)$ & Set of RL actions \\
      $R(s,a)$ & Reward for RL\\
      $w$ & Weights of the RL Reward function\\
      $\psi$, $\vartheta$, $\eta$, $\upsilon$ & Delay, security, reliability, accuracy of the event\\
      $\epsilon$ & Exploration probability\\
      $s'$,  $a'$ , & Next state and next action for RL\\
      $\alpha$, $\gamma$ & Learning rate and discount factor for RL\\
      $Q_t(s, a)$ & Q-Value of state $s$ and action $a$ for RL \Bstrut \\
      \hline
    \end{tabular}
  \end{center}
\end{table}

\section{Q-learning Based Update Decision}\label{sec:solution}
The previous problem formulation requires optimizations, which even for small networks, and multiple versions of ML models, and under multiple constraints, can be prohibitively complex. We therefore propose a practical Update Agent implementation. We use RL to make decisions on updates when either: i) creating a model replica during system scaling, or, ii) processing a request from the queue. The former deploys a replica with either the most recent version of the ML model, or the initial version; the latter makes the decision based on whether the currently deployed replica should be used, or a newer version, if available, is to be deployed.

The RL model used here includes the commonly known key elements:  agent, state space $S$, action space $A$, system reward $R$, and environment. RL agent computes the delays, security, model accuracy and reliability values associated with function requests. RL model employs a Q-table to record, at each decision-making stage, the information required by the agent to make an informed choice. This includes the current state, the action to be taken, the associated reward, and the subsequent state.
The system state $S$ at time $t$ for the RL model is represented
by: resource availability, ML model id,  queue length, event type and a set indicating the types of updates available for the ML model at stake:

\begin{equation}
    S = \{ s \mid s = (Z_k, f_k, Q_k, e, \Omega) \}
    \label{eq:state}
\end{equation}

where set $Z_k = \zeta_{1,k}, ..., \zeta_{n,k}, ..., \zeta_{N,k}$ denotes the nodes availability in a binary manner, and $\zeta_{n,k}$  is calculated from $\zeta_{n,k} = CPU_{n,k} \land RAM_{n,k} \land Disk_{n,k}$. If enough resources are available for a deployment of ML model $k$ in node $n$ $CPU_{n,k}$ is equal 1; otherwise 0. The same principle applies for $RAM_{n,k}$ and $Disk_{n,k}$. Consequently, the value of $\zeta_{n,k}$ equals 1 if there are enough resources available for CPU, RAM and Disk to deploy the ML model. Therefore, the value of $\zeta_k$ indicates whether ML model $k$ can be deployed on node $n$.
The ML model identification $f_k$ is an integer variable that defines the type $k$ of the model.
$Q_k$ is an integer variable that defines the queue size for this ML model.
A binary variable $e$ is used to indicate the type of event that occurs in the system. In the case of an arrival event, the value of $e$ is set to zero, while in the case of a departure event, the value of $e$ is set to one.
The set, noted as $\Omega$ is a list of length four, in which each element is a binary value  $\Omega = [\omega_1, \omega_2, \omega_3, \omega_4]$. Each $\omega$ represents a distinct type of update that differentiates the two ML model versions under consideration: main version, subversion security, subversion reliability, or subversion accuracy updates.

The \textit{Update Manager} takes a binary set of actions $a(s)$, i.e., 

\begin{equation}
    a(s) = 
    \begin{cases} 
        1 & \text{if instance } m \text{ is updated} \\
        1 & \text{if new instance } n \text{ is created with newest version} \\
        0 & \text{otherwise}
    \end{cases}
    \label{eq:action}
\end{equation}

At each system state and following the execution of an action $a(s)$, the update decision agent receives a reward. 
The reward function $R(s,a)$ used in our update decision agent considers the total delay, in addition to the values of security, reliability and accuracy. The agent will select actions that minimize the delay while simultaneously optimizing security, reliability and accuracy. The reward is defined as follows:

\begin{equation}
    R(s,a) = - w_1 \psi + w_2 \vartheta + w_3 \eta + w_4 \upsilon
    \label{eq:reward}
\end{equation}
where $s$ is a state, $a$ is an action, $\psi$ is the total delay for the processed event, $\vartheta$ is the security value, $\eta$ is the reliability value and $\upsilon$ is the accuracy value of this event. 
\begin{algorithm}
\footnotesize 
	\caption{RL-based Update Decision Algorithm}
	\label{alg:UpdateDecision}
	\begin{algorithmic}[1]
		\State \textbf{Input: } events, $b_k$,  $\mu_k$, network state ($Z$, $Q$)
		\State \textbf{Initialization: }$\mathcal{Q}_t$, $\epsilon$, $\alpha$, $\gamma$, episodes, decay
		\For {each event}
        {
            \If{update available}
                \State $s$ $\leftarrow$ ($Z$, $Q$, $e$, $f$), \;\;\;   $a$ = UpdateDecision($s$) 
                \State $Z'$, $Q'$, $e'$, $f'$ $\leftarrow$ Find next state parameters
                \State $s'$ $\leftarrow$ ($\Delta'$, $Q'$, $e'$, $f'$)
                \State $\mathcal{Q}_t$ $\leftarrow$ UpdateQTable($s$, $a$, $s'$)
                \State $s$ $\leftarrow$ $s'$
            \Else
                \State pass
                \EndIf
            \If{$\epsilon$ $>$ $\epsilon_\text{min}$}
                \State $\epsilon$ $\leftarrow$ decay $\cdot \epsilon$
                \EndIf
        }
        \EndFor
				
	\end{algorithmic}
\end{algorithm}

Algorithm \ref{alg:UpdateDecision} shows how we select an action,
as defined in Eq. \ref{eq:action}. 
During warm-up, where previous actions were taken, or the RL model has not learned yet
to accurately choose actions, an exploration phase is needed. We
consider an $\epsilon$-greedy approach to explore the search space and
try new actions. We denote by $\epsilon$ the exploration probability,
where at each step the RL agent chooses randomly an action
with probability $\epsilon$ and uses the accumulated knowledge with
probability $1 - \epsilon$. As the simulation progresses, the exploration-exploitation factor declines linearly until it reaches a minimal value.
At each point of decision, the agent checks the
state of the environment from Eq. \ref{eq:state} to take an action
a using $UpdateDecision(s)$ function, which chooses
an available action with the highest q-value, considering the $\epsilon$-greedy approach.


\begin{table}[]
\centering
\resizebox{\linewidth}{!}{%
\begin{tabular}{|
>{\columncolor[HTML]{EFEFEF}}l |c|c|c|c|}
\hline
Node & \begin{tabular}[c]{@{}l@{}}CPU Capacity\\ (in CPU cores)\end{tabular} & \begin{tabular}[c]{@{}l@{}}RAM Capacity\\ (in GB)\end{tabular} & \begin{tabular}[c]{@{}l@{}}Disk Capacity\\ (in TB)\end{tabular} & \begin{tabular}[c]{@{}l@{}}Transmission time\\ to master\end{tabular} \\ \hline

1,2,3,4    & 16                                                                    & 16                                                             & 1                                                               & 0, 2.75, 7.25, 10.25                                                                 \\ \hline
\end{tabular}
}
\caption{Topology parameters analyzed.}
\label{table:topology}
\end{table}


\section{Numerical Evaluation}\label{sec:results}
\begin{table}[]
\resizebox{\linewidth}{!}{
\begin{tabular}{|l|l|l|l|l|l|}
\hline
\rowcolor[HTML]{EFEFEF} 
ML Model id                     & 1                              & 2                              & 3                              & 4          & 5                              \\ \hline
\rowcolor[HTML]{EFEFEF} Version &   \multicolumn{5}{c|}{0 to  10   }  \\ \hline
\rowcolor[HTML]{EFEFEF} 
Subversion                      & \multicolumn{5}{c|}{0 to $\sim$ 10000  }         \\ \hline
Avg service time                & \multicolumn{5}{c|}{ 10 }                 \\ \hline
Avg inter-arrival time                &  \multicolumn{5}{c|}{ 3.25 - 8  }           \\ \hline
CPU req                         & 1                              & 2                              & 3                              & 4          & 5                              \\ \hline
RAM req                         & 1                              & 1                              & 2                              & 2          & 6                              \\ \hline
Disk req                        &  \multicolumn{5}{c|}{0.01 }          \\ \hline
Spawn time                      &  \multicolumn{5}{c|}{10 }            \\ \hline
Reliability                     &   \multicolumn{5}{c|}{0.9 - 1 }      \\ \hline
Security                        &   0.6 - 1 & 0.6 - 1 & 0.6 - 1 & 0.7 - 1 & 0.6 - 1    \\ \hline
Accuracy                        &   0.5 - 1 & 0.5 - 1 & 0.5 - 1 & 0.7 - 1 & 0.7 - 1     \\ \hline
\end{tabular}
}
\caption{ML model parameters simulated.}
\label{table:funmodels}
\end{table}

We now numerically analyze the efficacy and performance of various update agent implementations in determining whether to proceed with an update to the newest version of an ML model, or to maintain operation of the existing stable ML version.  
To this end, we use an event based simulator with negative exponentially distributed arrivals and service times, and provide the results with 98\% confidence level. 
We use a first-fit algorithm for resource allocation of the ML model placement to workers. 
We applied a common, monitoring-based scaling algorithm for creating and removing replicas, akin to \cite{10437794} and \cite{10279357}.
This is a straightforward algorithm, commonly used in practice, which monitors and analyzes the worker load. In the event that the load exceeds a specified threshold, the creation of a new replica is be scheduled. If all requests within the queue are processed and the replica is in a state of idle mode, awaiting the arrival of a new request, it is deleted. Otherwise the monitoring-based scaling algorithm will not take any action.
In addition to the RL based Update Agent, we implement a few more practical approaches: i) \emph{always update}, ii) \emph{never update} and iii) \emph{select an update randomly}.
Our topology consists of 4 nodes arranged in a star topology. Node 1, situated at the center of the network, serves as the master node for the cluster and also functions as a worker node, see Table \ref{table:topology}. All nodes are equal, and have 16 CPU cores, 16 GB of RAM, and 1 TB of disk capacity, with the transmission delay between each node to the master node being 0, 2.75, 7.25, and 10.25 (ms), respectively. 

We simulate 5 distinct ML models, as outlined in Table \ref{table:funmodels}. Each ML model, when first deployed at the beginning of the simulation, will have version $0.0$ and have 10 main version updates, e.g 1.0, 2.0 etc. Between the main version updates the ML model will have $\sim10000$ subversion updates, e.g 0.1, 0.2, etc. 
To vary the system load we assume a fixed average service rate, that is the same for all ML models and execute the simulation with different average inter-arrival times.
For various ML models, we assume different CPU and RAM requirements. For the sake of simplicity, it will be assumed that these requirements will remain constant with any new releases of the same model.
Similarly, a fixed spawn time will be assumed for all ML models across all versions. 
Reliability, security and ML model accuracy will be initialized to a fixed value for each ML model and will be improved with new version releases.
In case of a main version update, security and accuracy of a ML model will be improved by an order of magnitude $\sim2\%$ and reliability $\sim0.5$, as with main version updates functionality gets often extended and major changes are applied.
A Subversion update will improve one of the values by an order of magnitude $<<0.1\%$, to imitate a small performance update or security bug fix.
For the reward function defined in Eq. (\ref{eq:reward}), the weights were chosen to be $w_1 = 1$ , $w_2 = 10$ , $w_3 = 10$ and  $w_4 = 10$. For Q-table update, we set $\alpha = 0.01$, $\gamma = 0.99$, the initial exploration
probability $\epsilon  = 1$ , this will decay with an $decay$ up to an minimal $\epsilon_\text{min} = 0.001$ which will be reached after half of the scheduled events have been processed. The RL settings are mostly used for training and proved a good performance during hyperparameter tuning.

\begin{figure*}
 \centering 
 \begin{subfigure}[t]{0.24\linewidth}
 \includegraphics[scale=0.28]{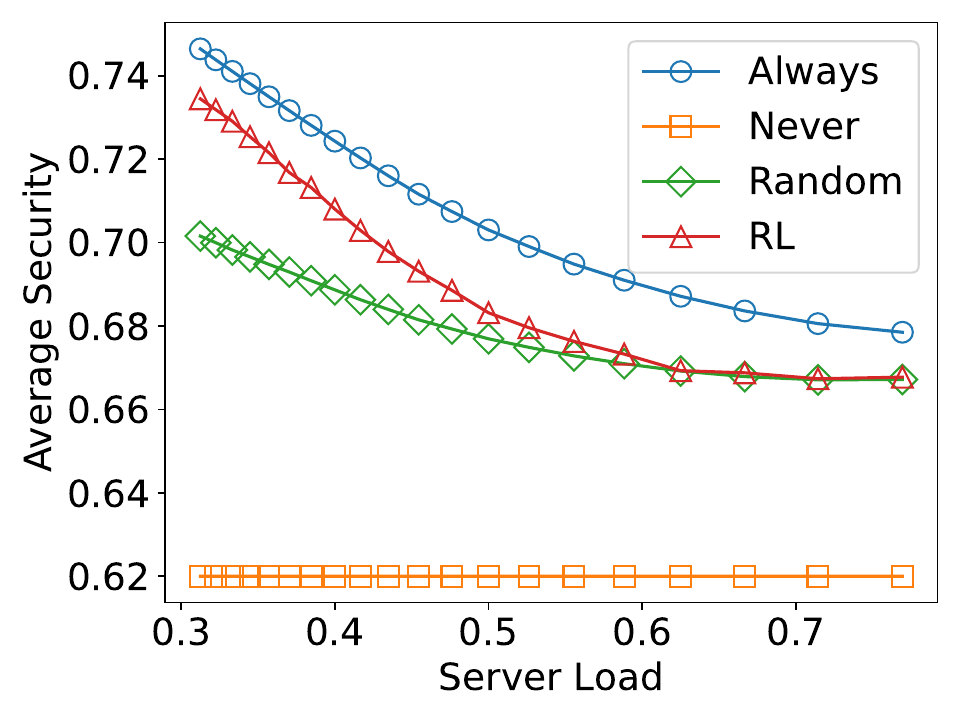}
  \caption{Average Security  }\label{subfig:mat1}
   \end{subfigure}
   \begin{subfigure}[t]{0.24\linewidth}
 \includegraphics[scale=0.28]{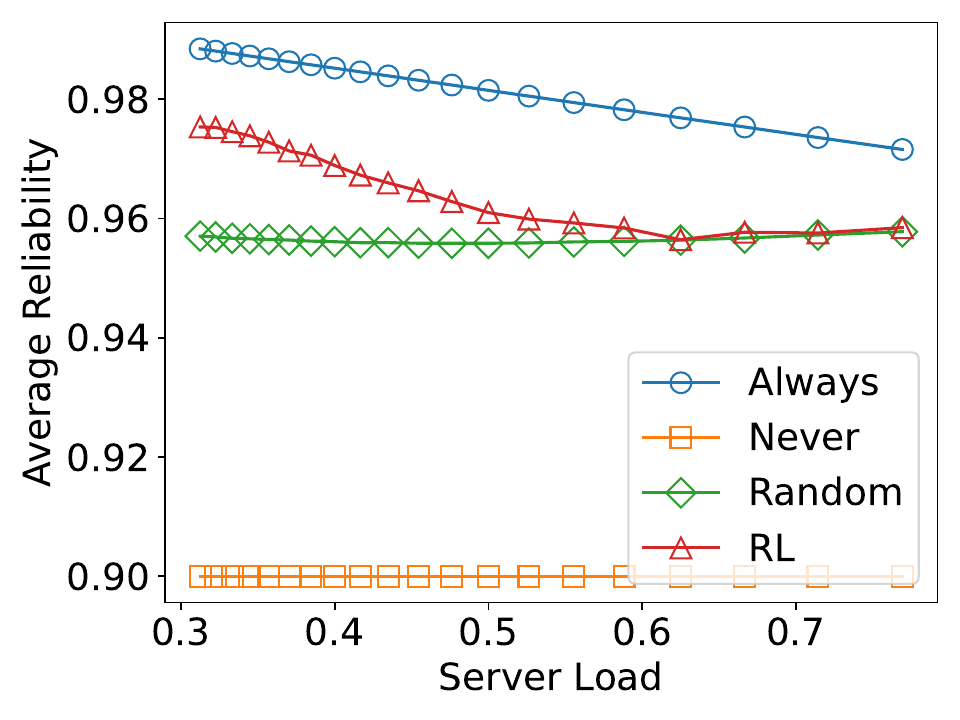}
 \caption{Average Reliability }\label{subfig:mat2}
   \end{subfigure}
 \begin{subfigure}[t]{0.24\linewidth}
 \includegraphics[scale=0.28]{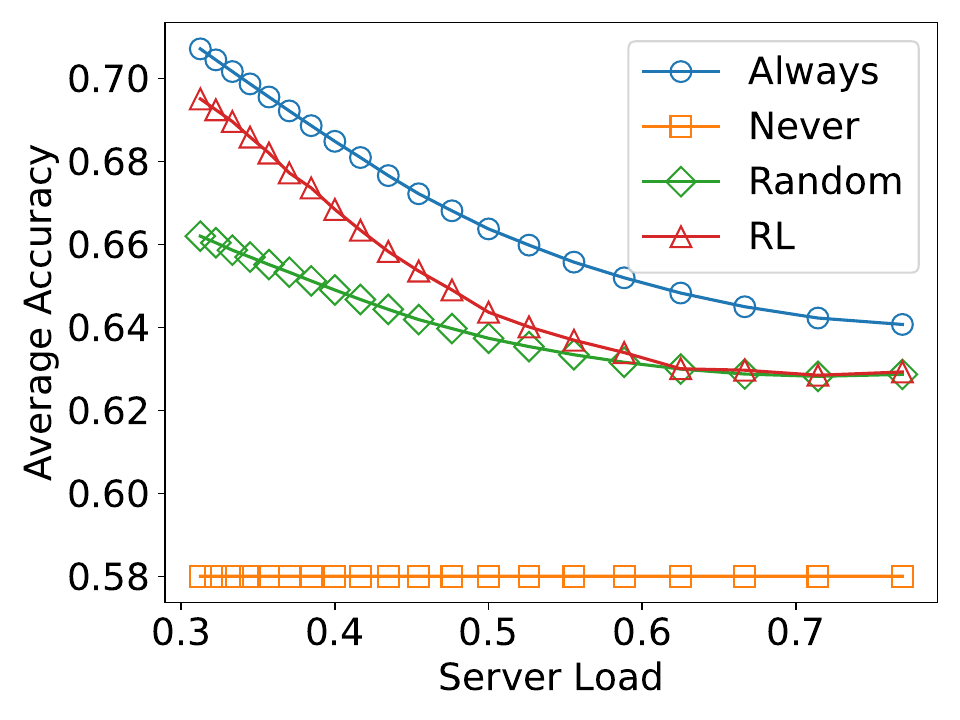}
 \caption{Average Accuracy  }\label{subfig:mat3}
   \end{subfigure}
    \begin{subfigure}[t]{0.24\linewidth}
 \includegraphics[scale=0.28]{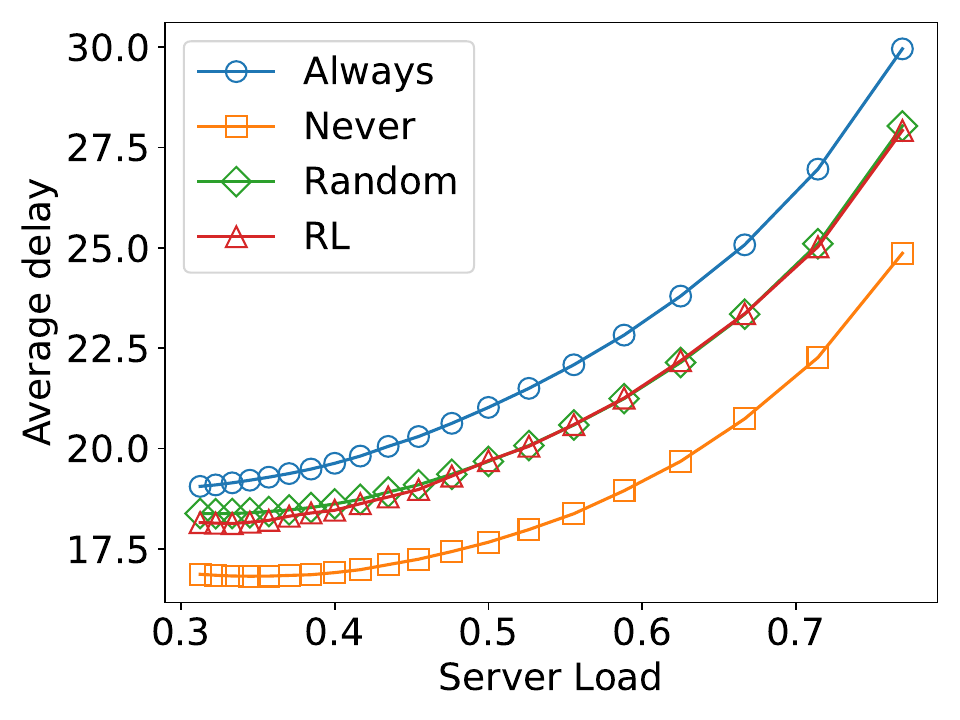}
 \caption{Average Delay  }\label{subfig:mat4}
   \end{subfigure}
 \caption{Performance metrics  with various loads of different update approaches. }
\label{fig:performance_metrics}
\vspace{-0.2 cm}
\end{figure*}

\begin{figure*}
 \centering 
 \begin{subfigure}[t]{0.24\linewidth}
 \includegraphics[scale=0.28]{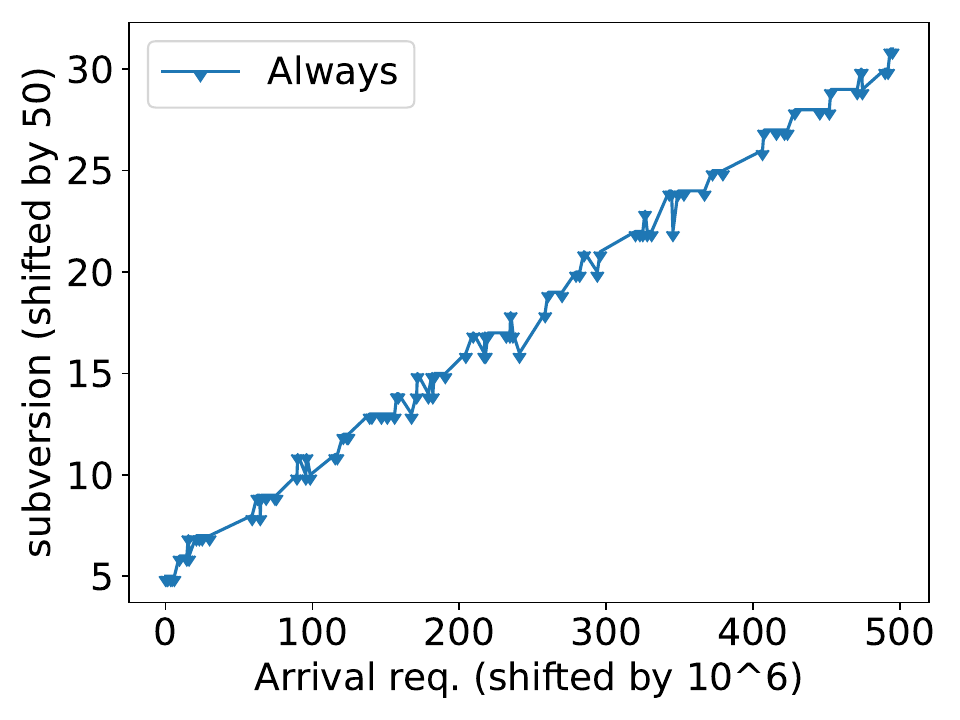}
  \caption{Always Update  }\label{subfig:sub1}
   \end{subfigure}
   \begin{subfigure}[t]{0.24\linewidth}
 \includegraphics[scale=0.28]{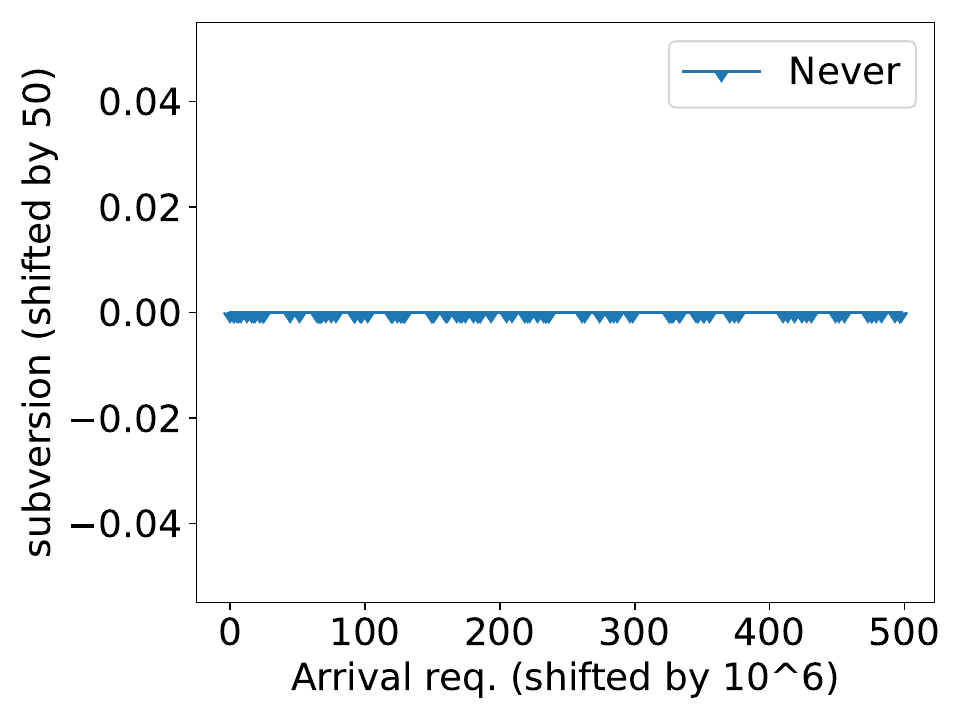}
 \caption{Never Update }\label{subfig:sub2}
   \end{subfigure}
 \begin{subfigure}[t]{0.24\linewidth}
 \includegraphics[scale=0.28]{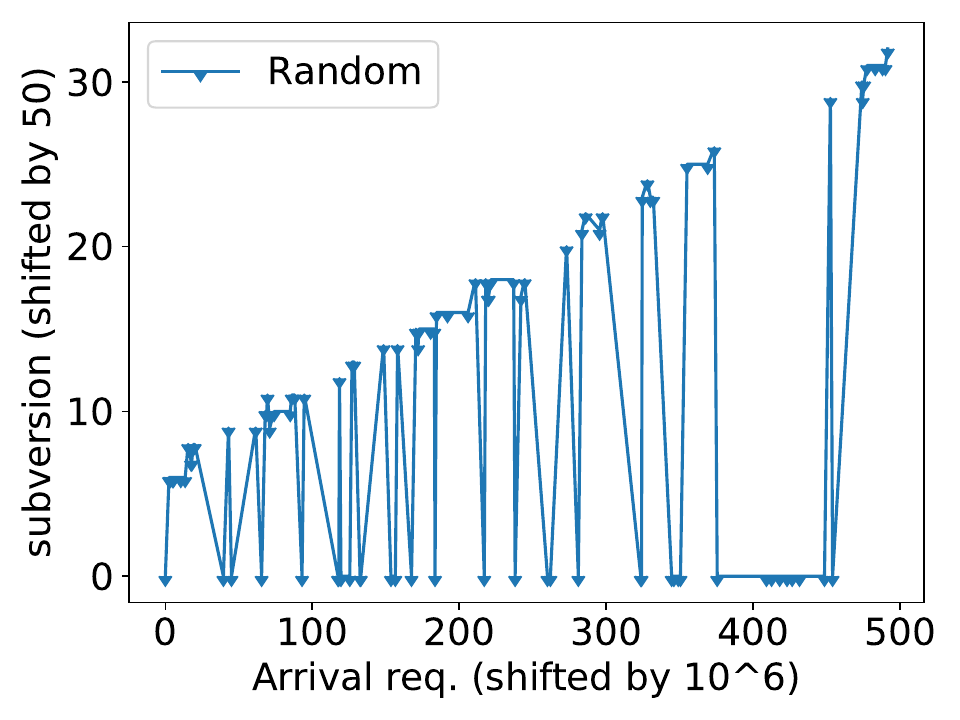}
 \caption{Random Update Decision  }\label{subfig:sub3}
   \end{subfigure}
    \begin{subfigure}[t]{0.24\linewidth}
 \includegraphics[scale=0.28]{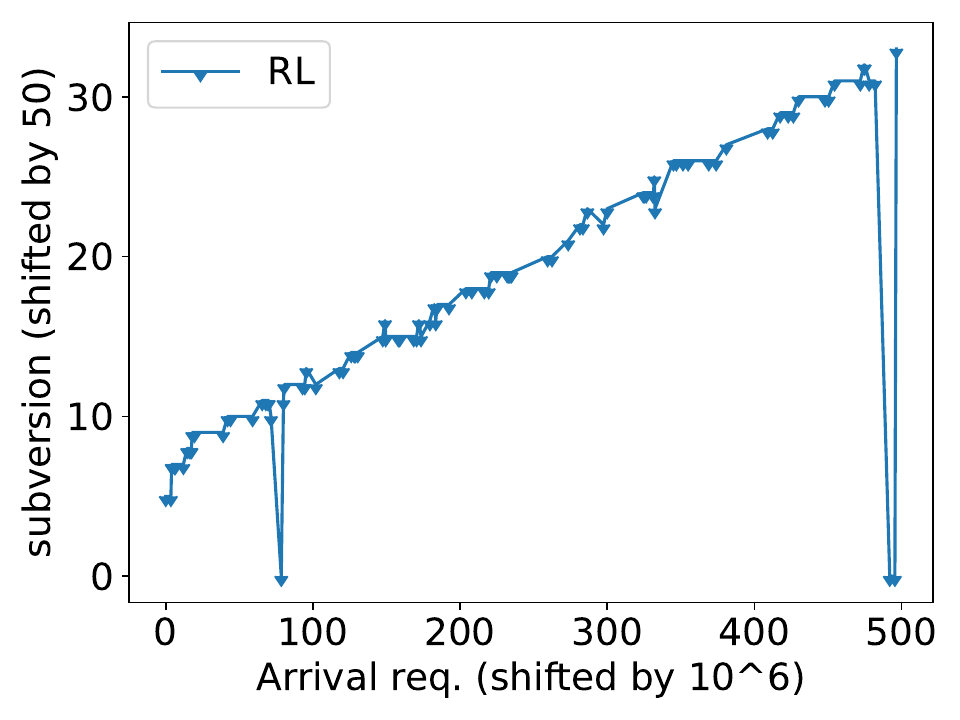}
 \caption{RL Update Decision  }\label{subfig:sub4}
   \end{subfigure}
 \caption{Used subversion over events of different update approaches for Load of 31.5\% in an interval for ML model 5}
\label{fig:fun5_subversion}
\vspace{-0.2 cm}
\end{figure*}

Fig. \ref{fig:performance_metrics} illustrates the performance metrics for various update approaches under different server load conditions using Eq. \ref{eq:load}, varying from 30\% up to a maximum value of about 80\%.
Figs. \ref{subfig:mat1}, \ref{subfig:mat2}, \ref{subfig:mat3} display  the average security ($\mathcal{O}_3$), average reliability ($\mathcal{O}_4$), and average accuracy ($\mathcal{O}_2$)  as a function of the system load. The performance achieved follow the same trend. As the server load increases, the parameters  for all update policies are observed to decrease. As expected, the method \emph{always update} is performing the best, as it consistently selects the most recent version with the best parameters (security, reliability, accuracy), i.e., with latest security updates.
Nevertheless, it decreases with increasing server load, because the higher the load the less updates received and processed by the system for the same number of requests. For low server load,  RL is the second-best approach after \emph{always update,} as it learns to enhance the parameter values.   For higher loads,  RL behaves similar to \emph{random update,} because when the server load is high, the system never scales-down replicas, which prevents the system from using old and stable versions and there is minor variation in the decision space. The performance of \emph{never update} remains consistent for all system loads, as the system maintains identical versions of all ML models. 

Figure \ref{subfig:mat4} shows the average delay for all update strategies. As expected, the delay increases with increasing load. \emph{Never update} performs the best, which avoids the additional spawn time periods. In order to update a deployed ML model,   we shut-down the existing replica and spawn a new one with the most recent ML model, which adds additional delays.  RL achieves a lower delay than the \emph{random} for lower server loads. For higher loads, it again exhibits behavior similar to \emph{random update}. As expected, \emph{always update} performs comparably worse, due to the additional delays at each possible update.

We also illustrate of the evolution of the deployed subversion for the various decision methods, see Figure \ref{fig:fun5_subversion}, because the deployed version and subversion determine the values that we aim to optimize ($\mathcal{O}_2$, $\mathcal{O}_3$, $\mathcal{O}_4$), while a change may impact the delay ($\mathcal{O}_1$). We show an interval of 500 arrival requests, beginning subsequent to the RL exploration phase after half of all arrival requests are processed, so displaying the interval from 1 million arrivals to 1,000,500. For better visibility the subversions are to be shifted by 50 in Figure \ref{fig:fun5_subversion}. As an example we are showcasing the deployed subversion for ML model 5 at a server load of 31.5\%. \emph{Always update} (Figure \ref{subfig:sub1}) shows the progressive increase in subversion over time. The minor regressions are due to a not-yet updated replicas for the same ML model, while other replicas have finished the update faster. \emph{Never update} (Figure \ref{subfig:sub2}) always deploys the oldest version and does not undergo any updates. \emph{Random update} (Figure \ref{subfig:sub3}) shows no clear trend. In the case of the decision to either update a replica or retain the existing version a more step-like function can be observed. When a new replica  is created, the option of selecting a replica with subversion 0 is made randomly, which explains the behavior of the curve.  Finally, RL strategy (Figure \ref{subfig:sub4}) illustrates a similar step-like behavior as \emph{random update} trajectory, not updating at every possible time point, while less often deploying the oldest most stable version of the ML model. 

 \section{Conclusion}\label{sec:conclusion}
In this paper, we addressed the problem of effective automation of ML model versioning in constrained edge networks. We compared a RL based approach to other update common strategies. 
The Update Agent, when implemented with reinfocement learning, showed a notable and practically relevant trend. A replica was more often updated in instances where multiple subversion updates were available, rather than deploying the most recent replica for each minor update. The simulation results demonstrate that the quality of RL-based decision depends on the server load. For less load, RL achieves comparable results to the \emph{always update} strategy for ML model specific parameters, while achieving better performance. Perhaps most importantly, our study showed important limitations and thus directions for further research. We assumed that subversion updates always improve (linearly or exponentially), which is practically not always the case. We did not take the stability as a values for the optimization problem into consideration yet, which also is premium in real world systems. Also, we did not take into account any downgrading of a version. These and many other interesting findings are subject of future research.

\section*{Acknowledgment}
This work was partially supported by the EU project MANOLO under GA no. 101135782.

\bibliographystyle{IEEEtran}
\bibliography{mybib}

%

\end{document}